\documentclass[twocolumn,showpacs,preprintnumbers,amsmath,amssymb]{revtex4}

\usepackage{graphicx}
\usepackage{dcolumn}
\usepackage{bm}

\begin{document}

\title{A Non-equilibrium STM model for Kondo Resonance on surface}
\author{Wei Fan}
\email{fan@theory.issp.ac.cn}
 \affiliation{ Key Laboratory of Materials Physics, Institute of Solid State Physics, Chinese
 Academy of Sciences, 230031-Hefei, People's Republic of China}

\begin{abstract}
 Based on a no-equilibrium STM model, we study Kondo resonance on a surface by self-consistent
 calculations. The shapes of tunneling spectra are dependent on the energy range of tunneling
 electrons. Our results show that both energy-cutoff and energy-window of tunneling electrons
 have significant influence on the shapes of tunneling spectra. If no energy-cutoff is used,
 the Kondo resonances in tunneling spectrum are peaks with the same shapes in the density of state
 of absorbed magnetic atoms. This is just the prediction of Tersoff theory. If we use an energy
 cutoff to remove high-energy electrons, a dip structure will modulate the Kondo resonance peak
 in the tunneling spectrum. The real shape of Kondo peak is the mixing of the peak and dip,
 the so-called Fano line shape. The method of self-consistent non-equilibrium matrix Green function
 is discussed in details.

\end{abstract}

\pacs{68.37.Ef, 73.23.-b, 72.10.-d, 73.20.-r}

\maketitle

\section{\label{sect1} Introduction }

 Scanning Tunneling Microscope (STM) has been widely used to study atomic
 and electronic structures of surface and atoms absorbed on it.
 A currently interesting topic is Kondo effect of magnetic atoms on non-magnetic metal
 surface. The interest along this direction is motivated by recent advance in the construction
 of novel nano-structure device on surface. STM has been used to probe the Kondo resonance
 of a single magnetic atom.\cite{Madhavan2,Nagaoka1,Knorr1,Schneider1,Jamneala1,Li1}
 The magnetic cobalt phthalocyanine molecule absorbed on Au(111) surface shows Kondo effect
 by having cut away eight hydrogen from the molecule.~\cite{Zhao1} Kondo resonance haven been
 found in other magnetic molecules absorbed on metal surfaces such as Co(CO)$_{n}$
 molecules on Cu(100) surface~\cite{Wahl1}. Additionally, STM has also been used to
 study the formation of Kondo molecule.\cite{Madhavan1,Kudasov1} The most interesting
 phenomenon is so called Quantum Mirage~\cite{Manoharan1,Fiete1} due to the refocus of
 Kondo resonance on surface.

 Generally STM tunneling spectrum isn't exactly correspondent to the atomic and electronic
 structure of the surface. To obtain the realistic atomic and electronic structure from STM
 tunneling spectrum, it is valuable to establish a believable STM theoretical model that
 is related the realistic atomic structure or electronic structure to the STM tunneling spectrum.
 Based on the Tersoff alike theory,\cite{Tersoff1,Tersoff2,Bardeen1,Bracher1} the differential
 conductance of electronic tunneling in STM is proportional to the Local Density of State(LDOS) of
 surface state. The resonance peak at zero-bias voltage in dI/dV curve is generally the mixing of Kondo
 resonance with other resonance such as d-resonance from 3d magnetic atom.\cite{Nagaoka1}
 The STM models based on Anderson Model have successfully explained the Fano shape of Kondo
 resonance using the equilibrium many-body theory~\cite{Schiller1,Lin1,Luo1,Ujsaghy1,Merino1}
 and non-equilibrium theory.\cite{Plihal1}

 Quantum tunneling of electrons in STM device is out of equilibrium in nature. In this paper,
 using a non-equilibrium self-consistent method, we calculate the STM tunneling spectra for magnetic
 atoms on metal surface. Kondo effect can be solved using analytical or numerical methods,
 such as,  the Non-Crossing Approximations (NCA) and other Larger N methods are suited for the
 infinite U Anderson model.\cite{Hewson1,Coleman1,Bickers1} The numerical renormalization group
 (NRG) and the exact Bethe Ansatz need to linearize the dispersion relation near Fermi energy
 after having simplified into one-dimension model.\cite{Hewson1} The method used in this paper is
 similar to the iterative perturbation theory (IPT).\cite{Kajueter1,Schonhammer1} The big difference is
 that in this work the Kondo resonance peak is decoupled from other parts of the Green
 function. However for IPT, the Kondo resonance is imported into the Green function of magnetic atom by
 interpolating $\Sigma_{0}^{2}$ self-energy into an appropriate self-energy ansatz.
 The decoupled scheme in this work can be obtained by more detail theoretical analysis such as in
 reference.\cite{Gunnarsson1}

 As a company of Kondo resonance, the so-called zero-bias anomaly has contribution to the differential
 conductance near zero-bias voltage. The zero-bias voltage anomaly may have different origins such as
 the orthogonality catastrophe~\cite{Anderson2} and  the co-tunneling mechanism~\cite{Franceschi1,Weymann1}.
 The electrons have no enough energies for most of many-body events at small bias voltage.
 The zero-bias anomaly appears as a local minimum in dI/dV curves near zero-bias voltage.

 Experimentally Kondo resonance in dI/dV curves are clear dip
 structures for single absorbed magnetic atoms~\cite{Madhavan2,Nagaoka1,Knorr1,Schneider1}
 but peak structures for absorbed magnetic molecules~\cite{Zhao1,Wahl1}.
 In this paper, the signals of Kondo resonance from the dI/dV curves are Fano-line
 shapes and are the mixing of Kondo resonance and dip structure due to zero-bias anomaly.
 This is well consistent with experimental observations. The zero-bias anomaly generally
 appears as a dip structure near zero-bias and is frequently covered by background noise.
 If other resonance peak appears at zero bias, the zero-bias anomaly can shows itself by
 modulating the resonance peak. This is why Kondo resonance generally appears company with
 a local minimum in dI/dV curves. If we properly reduce the back ground noise, the dip
 structure also can be observed in the tunneling spectra. The shapes of tunneling
 spectra are dependent on the energy-range of tuneling electrons. The paper is organized as following,
 after brief introduction we present the theoretical STM model in section two. In the section three,
 we present the non-equilibrium self-consistent method in detail. Our main results are presented in
 section four. Two appendixes are at the end of the paper which include the brief introduction of the
 matrix representation of non-equilibrium Green function and the derivation of the tunneling-current
 formulations.
 \begin{figure}\scalebox{0.8}{\includegraphics{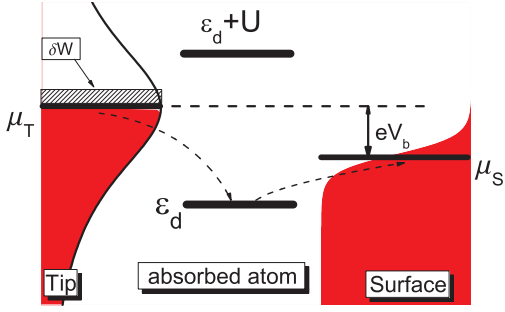}}
 \caption{\label{fig1} The energy-level distribution of the STM device. Two horizontal lines labelled with
 $\mu_{_{T}}$ and $\mu_{_{S}}$ illustrate the chemical potentials of the tip and the surface.
 The $\mu_{_{S}}$ is the chemical potential for the surface where the Kondo resonance happens.
 The red areas show the states of the tip and surface, which are filled with electrons. The
 chemical potential $\mu_{_{S}}$ of the surface is fixed to zero. The difference of the chemical
 potentials for the tip and the surface is induced by the bias voltage V$_{b}$ applied to the surface.
 Two thick bold lines present the splitting energy-levels for magnetic atom due to the on-site
 coulomb interaction. Two curved dash lines show a cotunneling event in the STM device.}
 \end{figure}
\section{\label{sect2} Theoretical STM Model}

 The Non-equilibrium Green-function method (Schwinger-Keldysh Contour)
 ~\cite{Schwinger1,Keldysh1,Mahan1,Zhou1,Kamenev1} has been used successfully to study
 the mesoscopic quantum transport of quantum dots and others mesoscopic quantum
 device.\cite{Caroli1,Meir1,Wingreen1,Jauho1,Baigeng1,Anantram1} The model Hamiltonian is splitting two
 parts $H =H_{0}+H_{T}$. $H_{T}$ is the tunneling Hamiltonian which
 expresses the electronic transport between different parts of the STM device such as tip,
 surface and absorbed atom shown in [Fig.~\ref{fig1}].

  \begin{eqnarray}
  \label{eq1}
        H_{0}&=& \sum_{k\sigma}\varepsilon_{pk\sigma}p_{_{k\sigma}}^{\dagger}p_{_{k\sigma}}
        +\sum_{n\sigma}\varepsilon_{tn\sigma}t_{_{n\sigma}}^{\dagger}t_{_{n\sigma}}
     \\ \nonumber
            &+& \sum_{\sigma}\varepsilon_{\sigma}d_{_{\sigma}}^{\dagger}d_{_{\sigma}}
        +Ud_{_{\uparrow}}^{\dagger}d_{_{\uparrow}}d_{_{\downarrow}}^{\dagger}d_{_{\downarrow}},
        \nonumber \\
  \label{eq2}
     H_{T}&=& \sum_{n\sigma\sigma'}V_{n\sigma\sigma'}t_{_{n\sigma}}^{\dagger}d_{_{\sigma'}}
         +\sum_{k\sigma\sigma'}A_{k\sigma\sigma'}p_{_{k\sigma}}^{\dagger}d_{_{\sigma'}}
      \\ \nonumber
      &+& \sum_{nk\sigma\sigma'}B_{nk\sigma\sigma'}p_{_{k\sigma}}^{\dagger}t_{_{n\sigma'}}
         +h.c.
 \end{eqnarray}
 \noindent $p^{\dagger},t^{\dagger},d^{\dagger}$ and $p,t,d$ are the electron creation and
 annihilation operators of surface, tip and absorbed atom respectively.
 $V_{n\sigma\sigma'}$, $B_{nk\sigma\sigma'}$ and $A_{k\sigma\sigma'}$ are
 the tunneling matrix elements or hybridizing  matrix elements between the tip and absorbed
 atom, the tip and surface, and the absorbed atom and surface. $k$ is the energy-level
 index of the surface, $n$ the STM tip and $\sigma$ is the spin index of electron.
 $\varepsilon_{pk\sigma}$, $\varepsilon_{tn\sigma}$ and
 $\varepsilon_{\sigma}$ are the energy levels of the surface, the tip and the
 absorbed atom respectively. $U$ is the on-site Coulomb energy of the absorbed atom.
 The hybridization Hamiltonian, the second term of $H_{T}$, has the same form as the
 tunneling Hamiltonian. Thus hybridization can treat the same foot as the tunneling events.

 The STM probes the tunneling current or the change of charges in the tip when a bias voltage
 applies to the surface. The operator of electron number at the tip is written as
 $N_{T}= \sum_{n\sigma}t_{_{n\sigma}}^{\dagger}t_{_{n\sigma}}$ and the electronic
 current is its time-derivative of
 $J_{T}= -e\frac{dN_{T}}{dt} = \frac{ie}{\hbar}[ N_{T}(t),H]$.
 After having been calculated the commutations the current is expressed as

 \begin{eqnarray}
 \label{eq3}
 \langle J_{T} \rangle= \frac{2e}{\hbar}Re(
                        \sum_{n\sigma\sigma'} V_{_{n\sigma\sigma'}}
                        \langle t_{_{n\sigma}}^{\dagger}d_{_{\sigma'}}
                        \rangle^{<}
                       +\sum_{nk\sigma\sigma'} B_{_{nk\sigma\sigma'}}
                        \langle t_{_{n\sigma}}^{\dagger}p_{_{k\sigma'}}
                        \rangle^{<}),
 \end{eqnarray}

 \noindent where the lesser Green function $G^{<}(t,t') = i<A(t')^{\dagger}B(t)>$. The similar
 formula has appeared in Ref.\cite{Plihal1} but replaced the real part with
 imaginary part. Two formulas are the same if having considered without image unit {\it i}
 in the definition of lesser Green function in that paper. In the appendix B, we derive bellow
 current formula using the matrix non-equilibrium Green functions.
 \begin{eqnarray}
 \label{eq4}
 \langle J_{T} \rangle &=& J_{1}+J_{2} \\ \nonumber
       J_{1} &=& \frac{-2e}{h}\int_{c} d\omega
                            Re Tr[G^{ct(r)}(\omega)\Sigma^{ct(<)}(\omega)] \\ \nonumber
       J_{2} &=& \frac{-2e}{h}\int_{c} d\omega
                            Re Tr[G^{ct(<)}(\omega)\Sigma^{ct(a)}(\omega)].
 \end{eqnarray}
 \noindent where {\it Tr} presents the summation for all energy-level and spin
 indexes $\sum_{nn'\sigma\sigma'}$.
 The superscript $c$ indicates that Green functions are matrixes defined in the appendix A.
 The $\tau$, $\tilde\tau$, $<$ and $>$ indicate the four elements of the matrix Green
 function Eq.~(\ref{eq1a}) and the {\it r} and {\it a} are the retard and advance Green
 function defined in Eq.~(\ref{eq4a}). The current J$_{1}$ flows into the tip
 from the surface and absorbed atom, and J$_{2}$ flow out of the tip to the
 surface and absorbed atom. The limitations of integral of the current
 is dependent on the problem we will consider and can choose Eq.~(\ref{eq12a}) or Eq.~(\ref{eq12b}).
 Near zero bias we need an energy cutoff and Eq.~(\ref{eq12b}) is a good choice.

\section{\label{sect3}  The Methods of Numerical Calculation}

 The non-equilibrium Green function is defined on a contour with two branches: the positive branch
 from negative infinite on time axis to positive infinite and the negative branch from positive
 infinite return to negative infinite. The contour green functions can be replaced by a matrix Greens
 functions. The matrix Green can be calculated using the standard perturbed methods in the same manner
 as the zero-temperature Green function in the standard textbook.\cite{Mahan1} In Appendix A, we present
 the main formulas and their computational methods.

  \begin{figure}\scalebox{0.7}{\includegraphics{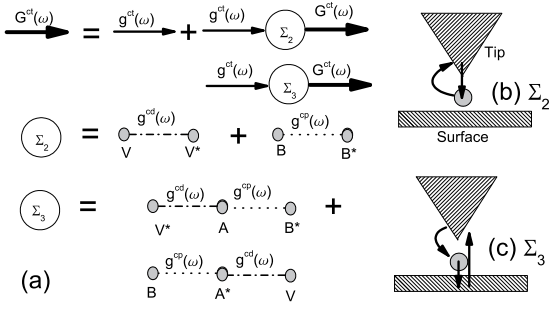}}
 \caption{\label{fig2} (a) Feynman diagram representation of Dyson equation for the
 tip. The Dyson equations for absorbed magnetic atom and surface are similar. (b) The physical
 event for the first term of the  tip self-energy $\Sigma_{2}$ includes two tunneling events.
 The electron jumps to the absorbed atom and returns to the tip. (c) The physical events for
 the first term of the tip-energy $\Sigma_{3}$ includes a cotunneling event from the tip to
 the surface via the absorbed atom and a single tunneling to return to the tip from the surface.}
 \end{figure}

 \subsection{The key formulas for self-consistent calculation}

 We assume the Hamilton H$_{0}$ can be solved exactly
 and the tunneling Hamilton treats as perturbations. The self-consistent calculations are used to
 obtain the full Green functions. At first, we solve the Hamilton H$_{0}$ and the obtained Green
 functions are used as the free Green functions for self-consistent calculations. At the next
 step, the self energies are calculated using the Eq.~(\ref{eq10a}). Three new Green Functions are
 calculated using three Dyson equations for the tip, adatom and surface, which are
 expressed as

 \begin{eqnarray}
 \label{eq5}
    G^{ct}&=& g^{ct}(\omega)+g^{ct}(\omega)\Sigma^{ct}(\omega)G^{ct}(\omega)
 \\ \nonumber
    G^{cd}&=& g^{cd}(\omega)+g^{cd}(\omega)\Sigma^{cd}(\omega)G^{cd}(\omega)
 \\ \nonumber
    G^{cp}&=& g^{cp}(\omega)+g^{cp}(\omega)\Sigma^{cp}(\omega)G^{cp}(\omega),
 \end{eqnarray}
 \noindent or diagrammatically illustrated in Fig.\ref{fig2}. We replace the free Green functions in
 the self-energy formulas Eq.~(\ref{eq10a}) with new Green functions.

 \begin{eqnarray}
 \label{eq6}
         \Sigma_{\alpha\beta}^{cd}(\omega) &=& \sum_{n'n''\sigma'\sigma''}
                         V_{_{n'\sigma'\alpha}}V_{_{n''\sigma''\beta}}^{*}G_{_{n'n''\sigma'\sigma''}}^{ct}(\omega)
                        \\ \nonumber &+& \sum_{k'k''\sigma'\sigma''}
                         A_{_{k'\sigma'\alpha}}A_{_{k''\sigma''\beta}}^{*}G_{_{k'k''\sigma'\sigma''}}^{cp}(\omega)
                         + \Sigma^{cd}_{3\alpha\beta}(\omega) \\ \nonumber
        \Sigma_{kl\alpha\beta}^{cp}(\omega) &=& \sum_{n'n''\sigma'\sigma''}
                         B_{_{n'\sigma'k\alpha}}B_{_{n''\sigma''l\beta}}^{*}G_{_{n'n''\sigma'\sigma''}}^{ct}(\omega)
                      \\ \nonumber   &+& \sum_{\sigma'\sigma''}
              A_{_{k\alpha\sigma'}}A_{_{l\beta\sigma''}}^{*}G_{_{\sigma'\sigma''}}^{cd}(\omega)
              + \Sigma^{cp}_{3kl\alpha\beta}(\omega) \\ \nonumber
      \Sigma_{mn\alpha\beta}^{ct}(\omega) &=& \sum_{\alpha'\beta'}
                         V_{_{m\alpha\alpha'}}V_{_{n\beta\beta'}}^{*}G_{_{\alpha'\beta'}}^{cd}(\omega)
                         \\ \nonumber &+& \sum_{k'l'\alpha'\beta'}
                         B_{_{mk'\alpha\alpha'}}B_{_{nl'\beta\beta'}}^{*}G_{_{k'l'\alpha'\beta'}}^{cp}(\omega)
                         + \Sigma^{ct}_{3mn\alpha\beta}(\omega).
 \end{eqnarray}

 The Eq.~(\ref{eq6}) and three Dyson equations Eq.~(\ref{eq5}) form a self-consistent loop.
 In Fig.\ref{fig2} and Eq.~(\ref{eq11a}), the $\Sigma_{3}$ self-energies represent the cotunneling
 mechanism which will contribute part of the zero-bias anomaly in dI/dV curve.

 \subsection{The free Green functions for the surface and tip}

 At first, we calculate the
 equilibrium retarded and advanced Green functions as the free Green functions used in following
 non-equilibrium self-consistent calculations. The surface is treated as two-dimensional
 electronic gas. The density of state (DOS) of two-dimensional electronic gas
 is flat and featherless with 4.0 eV width centered at Fermion energy of the surface.
 The well-shaped sharp STM tip is constructed with smaller number of atoms. The tip can
 be modeled as an atomic cluster with discrete energy-levels. Because the electron number is
 limited, there exists a highest-occupied energy-level such as the HOMO (Highest Occupied Molecular
 Orbit) energy E$_{_{HOMO}}$ for molecular systems or Fermi-energy for metal, all
 states with energies less than the highest-occupied energy-level are occupied, and all states
 with energies larger than the energy-level are empty. The electrons with energies near the
 energy-level are important to the physical and chemical properties of atomic clusters.
 Because the tip is in the metal environment, we assume the tip is in metal state and has
 a sharp Fermi surface. The tip models as a single energy level identified as Fermi energy $\mu_{T}$.
 The free retarded Green function for the tip is written as
 $g^{rt}_{n\sigma}=\frac{1}{\omega-\varepsilon_{tn\sigma}+i\delta_{t}}$, where
 $\varepsilon_{tn\sigma}=\mu_{T}$. The DOS of the tip has the Lorentz shape and centers
 at its Fermi energy $\mu_{T}$ with width $\delta_{t}$.

 \subsection{The free Green functions for magnetic atom}

 The main term produced the Kondo effect is the spin-flipping term, that is the last term in the
 Hamilton $H_{0}$. We assume the Hamilton had been solved {\it exactly}. The free Green function of magnetic
 atom including the Kondo resonance is calculated using the Hartree-Fock approximation with
 U$^{2}$ and U$^{4}$ order self energy.\cite{Yamada1} The retarded Green function of the Anderson impurity is
 written as $G^{r}_{d\sigma}(\omega)=\frac{1}{\omega-\varepsilon^{d}_{\sigma}+i\Delta}$
 where  $\varepsilon^{d}_{\sigma}=\varepsilon^{d}+n_{\bar\sigma}U$.
 From the results of particle-hole symmetric Anderson model in reference,\cite{Yosida1,Yamada1}
 up to the $U^{2}$ and $U^{4}$ self energy, the G$_{K}(\omega)$ near zero frequency is written as

 \begin{eqnarray}
    \label{eq8}
    G_{K}(\omega) &=& \frac{1}{\omega-\Sigma_{2}(\omega)+i\Delta}
 \\ \nonumber
    \Sigma_{2}(\omega)&=&
 -\alpha_{R}\omega-i\alpha_{I}\Delta ((\frac{\omega}{\Delta})^{2}+(\frac{\pi T}{\Delta})^{2}).
 \end{eqnarray}

 \noindent where
 $\alpha_{R}=(3-\frac{\pi^{2}}{4})(\frac{U}{\pi\Delta})^{2}+0.0553(\frac{U}{\pi\Delta})^{4}$,
 $\alpha_{I}=\frac{1}{2}\left[(\frac{U}{\pi\Delta})^{2}+6(5-\frac{\pi}{2})(\frac{U}{\pi\Delta})^{4}\right]$
 and $\Delta=\pi\rho_{0}|A|^{2}$. We find
 from our calculations that it is enough up to $U^{2}$ term.

 At zero temperature, the height of the peak is exactly $\rho_{0}=\frac{1}{\pi\Delta}$.
 $\Sigma_{2}(\omega)$ satisfies the general Fermi-Liquid properties
 $Im\Sigma_{2}(0)=0$ and $\frac{dIm\Sigma_{2}(0)}{d\omega}=0$ at zero
 temperature.\cite{Luttinger} It is also a good approximation to asymmetric Anderson model in
 Kondo regime if the asymmetric Anderson model is still described approximately by Fermi liquid.
 The approximation has been used to study Kondo resonance on surface.\cite{Nagaoka1}
 The main temperature effect is introduced by Eq.~(\ref{eq8}) and other temperature effects are
 introduced by Fermi distribution function $n_{f}(\omega)$.

 The Green function of magnetic atom, similar to the reference,\cite{Ujsaghy1}  is written as
 \begin{equation}
   \label{eq7}
  g^{r}_{d\sigma}(\omega)=\hat{g}^{r}_{d\sigma}(\omega)+Z_{\sigma}G_{K}(\omega-\delta E),
 \end{equation}
 \noindent where $G_{K}(\omega)$ is Kondo resonance peak and $\delta E$  is the
 shift of Kondo peak away from the Fermi energy. $\delta E$ is generally very small about
 several meV. The choice of $Z_{\sigma}$, $(Z_{\uparrow},Z_{\downarrow})$=(1.0, 0.0),  meets
 the condition that the height of Kondo peak is equal to $\frac{1}{\pi\Delta}$ in the total
 density of state of magnetic atom. Instead of in reference~\cite{Ujsaghy1} where the Kondo
 resonance is Lorentz shape, in this work the
 treatment of Kondo resonance is improved with non-Lorentz shape.
 The Green function $\hat{g}^{r}_{\sigma}=\frac{1}{\omega-\varepsilon^{d}_{\sigma}+i\delta_{d}}$,
 where $\delta_{d}$ doesn't include the contribution from the hybridizations with surface
 state and tip and doesn't contribute to Kondo resonance. $\delta_{d}$ can be seen as the
 contributions from the bulk electrons~\cite{Ujsaghy1,Agam1} or other complex interactions
 between the absorbed magnetic atoms and the surface, and treated as an input parameter.
 The scheme Eq.~(\ref{eq8}) isn't simple decoupling because the width and height of the decoupling
 Kondo resonance are dependent on the energy-level width and splitting of magnetic atom. The Friedel
 sum-rule Eq.~(\ref{eq9}) also tights the Kondo resonance and other parts of Green function together
 in the following self-consistent calculations. At present, the scheme only used in
 the study of Anderson model in Kondo regime.

 \subsection{Summary of self-consistent calculations}

 Up to now, we have known the
 free retard Green functions for the tip, surface and magnetic atom. Based on
 Eq.~(\ref{eq5a}), we can transform the retard Greens to the Matrix Green functions as
 the free matrix Green functions for non-equilibrium self-consistent loops.
 In this work, the trial Green functions $G^{c}_{0}(\omega)$ for the self-consistent
 calculations are equal to their free Green functions $G^{c}_{0}(\omega)=g^{c}(\omega)$.
 In the self-consistent loop we calculate the occupation number

 \begin{equation}
    \label{eq9}
   n_{\sigma}=
  -\frac{1}{\pi}\int_{-\infty}^{+\infty}d\omega f(\omega-\mu_{S})ImG_{d\sigma}^{r}(\omega).
 \end{equation}
 \noindent the new position of energy-level of magnetic atom
 $\varepsilon^{d}_{\sigma}=\varepsilon^{d}+n_{\bar\sigma}U$ and the energy-level width
 $\Delta=-Im\Sigma^{d}(\omega=\mu_{S})$ which is the sum contributed from the hybridizations with surface
 state and tip. Thus the shape of Kondo resonance is also changed in each loop. If the changes of
 occupation numbers are smaller than 0.001, the occupation numbers, Green functions and self-energies
 converge simultaneously. The Friedel sum-rule Eq.~(\ref{eq9}) is satisfied and the self-consistent
 calculations stop.

 The spectral functions are obtained by formula
 $ A(\omega)=\frac{1}{2}(G^{<}(\omega)-G^{>}(\omega))$ and the density of State
 $\rho(\omega)=\frac{1}{\pi}A(\omega)$. The tunneling current is calculated using
 Eq.~(\ref{eq4}). The differential conductance is numerically calculated from the curve of the
 tunneling current via bias voltage. In our self-consistent perturbed calculations the basic
 equation $ G^{\tau}+G^{\tilde{\tau}}=G^{>}+G^{<}$ for the matrix Green function
 is only approximately satisfied with accuracy about from 0.001/eV to 0.01/eV.

\begin{table}
\caption{\label{tab1} The parameters for numerical calculations
(unit eV). The band width for surface state is 4.0 eV}
\begin{ruledtabular}
\begin{tabular}{lllllllll}
A & V & B & U & $\varepsilon_{d}$ & $\delta_{d}$ & $\delta_{t}$ \\
\hline  0.20 & 0.10 & 0.05 & 2.80 & -0.80 & 0.30 & 1.00
\end{tabular}
\end{ruledtabular}
\end{table}
 \begin{figure}\scalebox{0.8}{\includegraphics{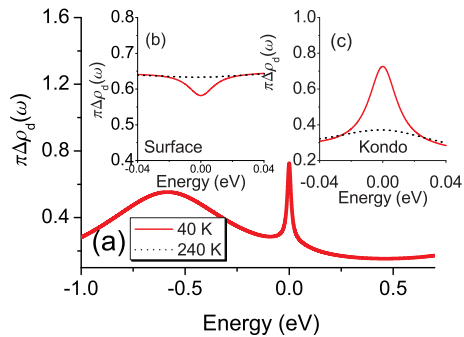}}
 \caption{\label{fig3} The density of state for magnetic atom (a,c) and the surface (b)
 with A=0.20 eV and U=2.8 eV. The broad peak in (a) is the wide energy level for magnetic atom; the small
 sharp peak is the Kondo resonance. The solid line is the density of state (DOS) at 40 K and dot line at
 240 K. The surface density of states and the detail structure of Kondo resonance are shown in the insert
 figure.}
 \end{figure}
 \section{\label{sect3} Results and Discussions}

  In this work we concentrate on the small bias voltage. It is convenient
  to drop the indexes of  $V_{n\sigma\sigma'}$, $B_{nk\sigma\sigma'}$ and $A_{k\sigma\sigma'}$
  and simplify to V, B and A without introducing confusion. The model parameters
  are collected in table~(\ref{tab1}). The parameters are close to Kondo regime
  ($U \gg \Delta < \varepsilon_{d}$) and close to the values for Co atom on Au(111) surface provided
  in reference.\cite{Ujsaghy1}. In Kondo regime, there is sharp Kondo resonance peak at Fermi energy
  besides the bare virtual resonance peak in the density of state of the absorbed magnetic atom.
 \begin{figure}\scalebox{0.7}{\includegraphics{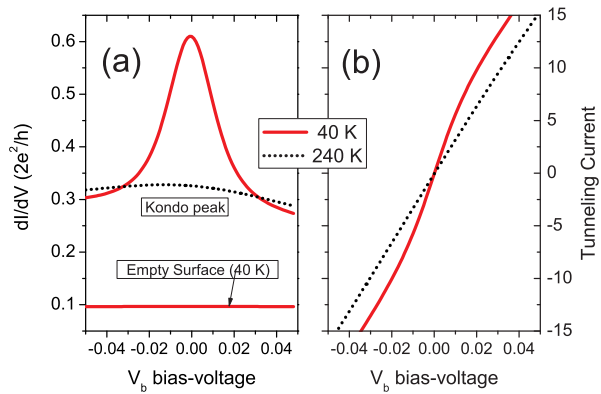}}
 \caption{\label{fig4} (a). The tunneling spectra of magnetic atom constant energy-window
 [-2.0 eV , 2.0 eV] at near zero-bias voltage at temperatures 40K and 240 K. Kondo peak
 becomes flat at 240K above Kondo temperature about 100 K. The tunneling spectrum of empty
 surface at 40 K also plots in this figure. (b) The tunneling current at 40 K and 240 K
 when the tip on top of the magnetic atom.}
 \end{figure}
 \subsection{Calculations of Density of state (DOS) and Differential conductance}

 The density of state of the absorbed magnetic atom and surface are shown in Fig.~\ref{fig3}.
 From this figure [Fig.~\ref{fig3} (a)], we can find one broad peak of energy
 level of the magnetic atom. The sharp peak at Fermi energy of the surface is well known Kondo
 resonance. The Fig.~\ref{fig3}(c) shows the flat effects of Kondo resonance, when temperature
 increasing above the Kondo temperature T$_{K}$. We estimate the Kondo temperature based on the
 width of Kondo Peak at zero temperature and find T$_{K}$=100 K. The Kondo temperature T$_{K}$
 also is estimated based on the equation $D\sqrt{J\rho}e^{-\frac{1}{J\rho}}$~\cite{Haldane1,Wilson1},
 where $\rho$ is the density of state of surface at Fermi energy,
 $J=\Delta / (\pi\rho)[1/|\varepsilon_{d}|+1/|\varepsilon_{d}+U|]$ obtained by Schrieffer-Wolff
 transformation~\cite{Schrieffer1} and $D$ is the effective band-width of surface state contributed
 to Kondo effects. We estimate the Kondo temperature T$_{K}$=123 K if 2$D$ = 4 eV.
 The 2$D$ is approximate to the energy windows in our calculations from -2 eV to 2 eV.
 The Fig.~\ref{fig3} (b) shows the density of state of the surface which is almost flat only there
 is a valley at the Fermi energy. The depression of surface DOS near Fermi energy is because some
 surface electrons at Fermi energy bind with the Anderson impurities and screen its local spin by
 forming spin-singlet binding state.

 The peak structures of Kondo resonance have been found in the tunneling spectra of
 cobalt atom embedded in larger molecules absorbed on metal surfaces.~\cite{Zhao1,Wahl1}.
 We calculate the differential conductance using the parameters in table~(\ref{tab1}) at
 temperature 40 K. The energy-range in the integral is from -2.0 eV to 2.0 eV,
 approximately from $-\infty$ to $+\infty$ in our calculations. We also calculate the
 tunneling spectrum of empty surface. From Fig.~\ref{fig4} (a) we can see that a peak
 appears at zero bias and it became flat at 240 K. It's just Knodo resonance in
 tunneling spectrum. If the tip move away from the magnetic atom to empty parts of the
 surface, the tunneling spectrum is flat. The peak structure of Kondo resonance and the
 flat spectrum for empty surface are celebrated because dI/dV curve is correspondent to
 the density of state and close to the prediction of the standard Tersoff's STM
 theory~\cite{Tersoff1,Tersoff2}.
 \begin{figure}\scalebox{0.7}{\includegraphics{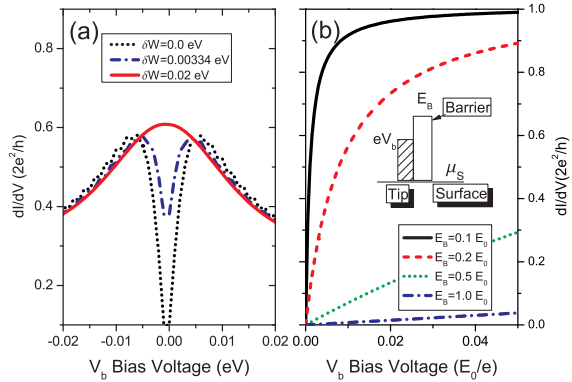}}
 \caption{\label{fig5} (a) The energy-window influences the shape of dI/dV curve near zero-bias.
 Three different energy-windows are used [-$(eV_{b}+\delta E)$,$(eV_{b}+\delta E)$]
 $\delta E$=0.0, 0.0034, 0.02 (eV). (b)The dI/dV curves for single electron transmits through
 a square barrier with height $E_{B}$. The sharp dip structures are shown for small
 barriers. The insert figure illustrates the simplified STM device without absorbed atoms
 on surface. $E_{0}=\frac{\hbar^{2}}{2ma^{2}}$ is a characteristic
 energy for the barrier}
 \end{figure}
\subsection{Influence of energy-window and energy cutoff on shape of resonance peak}

 The dip structures are more common in STM tunneling spectrum. The energy-window of tunneling
 electrons has important influence on the shapes of dI/dV curves. Several theoretical
 works~\cite{Merino1,Lin1} have shown that, in order to obtain correct tunneling spectra,
 the calculations of tunneling matrix need to use an energy-cutoff above Fermi energy
 to remove high-energy surface electrons. This is implicit that only part of electrons
 with proper energy-range can tunnel into the tip. At zero temperature, there exists clear
 energy-window with the Fermi energies of the tip and surface as its up-boundary and down-boundary.
 Only electrons with energies in the window can transmit to the other side of the barrier between
 the tip and surface and tunnel into the tip. The tip can't capture high-energy
 electrons although they have already transmitted the barrier.

 At finite temperature, the energy distribution of electron becomes smoothly across the Fermi energy.
 We may add a skip-energy $\delta W$ to define energy-window at finite temperature to include more
 electrons and cutoff electrons with higher energy. We alter the energy-window to
 [-(eV$_{b}+\delta W$),(eV$_{b}+\delta W$)] and calculate the dI/dV curve again.
 The width of energy-window changes with the bias voltage. In fact cutoff effect of Fermi energy has
 already included in our current formula by lesser Green functions and self energies for
 non-interaction case. After self-consistent calculations the electronic interactions of different
 parts are introduced. The cutoff effect of lesser Green functions becomes weak even at very low
 temperature and Fermi distribution still allows high-energy electrons above Fermi energy. We use
 the energy cutoff eV$_{b}+\delta W$ to neglect electrons with higher energies.
 The integral limits are symmetric around Fermi energy. From Fig.~\ref{fig5} (a), we can  see that
 the dip structures appear when $\delta W < 0.02$ (eV).
 For $\delta W=k_{_{B}}T=0.00334 (eV) <0.02 (eV)$, there is a sharp dip structure.
 We can also find from Fig.~\ref{fig5} (a) that energy-window effect is only effective near zero
 bias. If the minimum value of effective energy-window is larger than 0.02 (eV), the influence
 of energy-window can be ignored.

 Both the energy-window and energy-cutoff effects together determine the local minimum of dI/dV
 curve at zero bias. We must analyze them more deeply. At large bias voltage $V_{b}$ the
 energies of most tunneling electrons are in energy-range from 0 eV to eV$_{b}$. The number of
 electrons with high-energies is relative small. At very small bias voltage, the number of electrons
 within the energy-range is deceased to very small value. At this time, if we remove high-energy
 electrons the total difference conductance will has a large decrease. Only $\delta W$ less
 than 0.02 eV, in another word, more high-energy electrons are removed the local minimum near
 zero-bias is more prominent.

 We study a simplified STM model, that is, electrons
 transmits though a single barrier which is vacuum region between the tip and surface.
 From the insert figure of Fig.~\ref{fig5} (b), the height of the barrier is $E_{B}$ and the energy
 of incident electron is $E$. By standard quantum mechanic calculation,
 the transmitting coefficient is
 \begin{equation}
 T(E)=
 \frac{1}{\frac{E_{B}^{2}}{E(E_{B}-E)}\sinh(\sqrt{\frac{2ma^{2}}{\hbar^{2}}(E_{B}-E)})+1}.
 \end{equation}
 \noindent for $E<E_{B}$ where $a$ is the width of the square barrier. The
 differential conductance is calculated by the Landauer-B\"uttiker type formula
 $G=\frac{2e^{2}}{h}T(eV_{b})$.  From the Fig.~\ref{fig5} (b). We find the sharp dip
 structure for small barrier. The transmission coefficients increase with energies of
 tunneling electrons. Compared with high-energy electrons, near zero bias the energies of
 tunneling electrons are small and the transmission coefficients are also small.

 The dI/dV curves obtained from our model have similar shapes to the dI/dV curves for the single
 barrier model is indicative that the dip structures are because the energies of tunneling near low
 bias locate in classic forbidden regime of the vacuum barrier between the tip and surface.
 We remove the high-energy electrons by introducing energy-cutoff is equivalent that we restrict
 the energies of tunneling electrons below the effective barrier and within classic
 forbidden regime near zero bias. The number of high-energy electrons is small, but their transmission
 coefficient is large. They will have important influence on differential conductance because if
 at low bias most of electrons have low energies and small transmission coefficient.
 Although STM device has simple structure, many things are unclear such as the shape and
 size of tip. Generally the tip can't capture all tunneling electrons. At low bias, most
 of electrons locate in low-energy regime. The high-energy electrons and low-energy electrons are in
 different energy-scale so high-energy events are irrelevant to low-energy events.
 This is the reason why we introduce energy-cutoff to remove high-energy electron.
 The realistic STM devices are more complex. We must known the exact information of surface
 state. Of course, the directions of velocity of tunneling electrons also determine whether
 the tip can capture these tunneling electrons. Thus we suggest that, to more close to realistic
 STM, we should add a captured probability factor $W(\omega)$ into the current formula, and it can
 be written as $J=\frac{2e}{h}\int W(\omega)T(\omega)d\omega$, where $T(\omega)$ is
 transmission coefficient. In this paper, $W(\omega)=\theta(\omega-\mu)\theta(\mu+eV_{b}-\omega)$.
 This is just equivalent to the introduction of energy-window in our calculations.
 This is also equivalent to limit all tunneling matrixes with non-zero values only in this
 energy-window just like in reference.~\cite{Merino1,Lin1} Besides energy, $W(...)$ could include
 position, momentum and any other more complex properties of STM device.

 Based on our results, the dip structures in dI/dV curve near
 zero-bias are induced by the tunneling among the tip, absorbed atom and surface because the
 energies of tunneling electrons are smaller than heights of effective barriers of corresponding
 tunneling process. Not only most of many-body events but also subsequent single tunneling events
 are suppressed. The electronic tunneling events include subsequent single tunneling, co-tunneling
 and more complex tunneling process. If the background noises are reduced by proper method,
 the dip structure can be found in the dI/dV curve. If there are magnetic atoms on surface,
 Kondo resonance will enhance the differential conductance near zero bias, the dip
 structure has chance to be observed by modulating the shape of Kondo resonance.
 Thus the most of Kondo resonance observed in STM tunneling spectra are modulated by a local minimum.
 \begin{figure} \scalebox{0.8}{\includegraphics{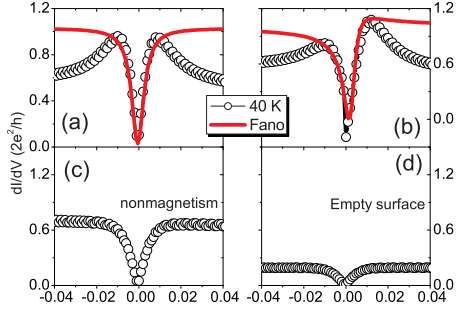}}
 \caption{ \label{fig6}The differential Conductance vs bias Voltage at 40 K with A=0.20 eV,
 U=2.8 eV. The solid lines with circles illustrate the results at 40 K. The Kondo temperatures
 T$_{K}$ are about 100 K. The solid lines indicate the Fano shapes fitting the dI/dV curves
 at 40 K. The figure (b) is the asymmetric resonance when the Kondo peak shifts away the Fermi
 energy about 0.005 eV. (c) the dI/dV curve for nonmagnetic impurity and (d) for empty surface.}
 \end{figure}
 \subsection{Calculations of differential conductance with proper energy-cutoff}

 In this section  we will study STM tunneling spectrum for the same parameters as in section
 A but we use a energy-cutoff from 0 eV to eV$_{b}$ (see Fig.~\ref{fig1}). The
 Eq.~(\ref{eq12b}) is well common used formula for the calculations of electronic transport
 near zero temperature. The results are consistent
 with most of STM tunneling spectrum of magnetic atoms on metal surface. The differential
 conductance (dI/dV) is plotted in Fig.\ref{fig6}(a). After having fitted the dI/dV curves
 at 40 K with Fano shape $\frac{(q+\varepsilon)^{2}}{1+\varepsilon^{2}}$ where
 $\varepsilon=\frac{\omega-\varepsilon_{q}}{k_{B}T_{q}}$, we get Fano parameter q=0.0 for
 (a) which is good fitting with Fano line-shapes. The symmetric resonant peak has been
 found in the STM tunneling spectra of Co/Au(111) and Co/Ag(111)~\cite{Schneider1}.
 Experimentally, some STM tunneling spectra of magnetic atoms on fcc(111) surface of
 novel metals are asymmetric dips~\cite{Madhavan2,Knorr1} or peaks.\cite{Nagaoka1}.
 We shift the position of Kondo resonance about $\delta E$ = 0.005 eV away from the Fermi
 energy. We obtain asymmetric resonance peak very close the shape observed in
 experiments~\cite{Madhavan2,Knorr1} and $q$ values by fitted with the Fano-line-shape is
 0.30 for Fig.\ref{fig6}(b).

 In the nonmagnetic calculation the parameters are the same but $U=0$. The clear dip structure is
 found from the dI/dV curve (Fig.\ref{fig6}(c)). For the empty surface without absorbed magnetic
 atoms the dI/dV curve has also a dip structure near zero-bias. There is no dip structure in
 the density of the state of the nonmagnetic impurity and empty surface at the Fermi energies.
 Thus dip in dI/dV curve is the tunneling phenomenon such as single tunneling and co-tunneling
 events. The magnetic impurity and even nonmagnetic can enhance the dip structure.
 The resonant structures near zero bias for nonmagnetic impurities are found in other non-equilibrium
 calculation such as in reference.~\cite{Plihal1}
  \begin{figure}\scalebox{0.8}{\includegraphics{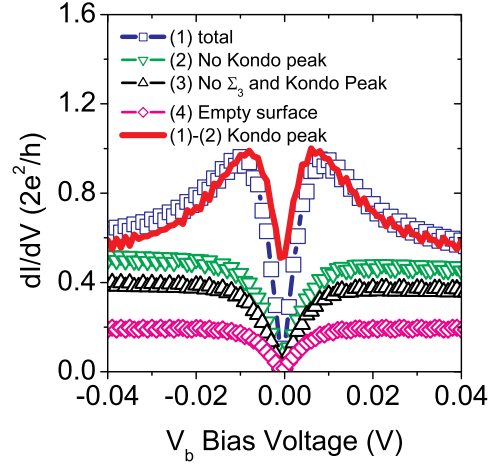}}
 \caption{\label{fig7}  (1) Open squares show the dI/dV curve for the full calculation including
 Kondo peak and (2) solid line with open down-triangles show the dI/dV curve without Kondo
 resonance ( Eq.~(\ref{eq8})) in self-consistent calculations.
 The (3) up-triangles for without both $\Sigma_{3}$ self-energies and Kondo peak, and the (4)
 diamonds  for empty surface are also plotted in this figure. The contribution of Kondo resonance
 to dI/dV curve (bold thick-solid line) is obtained by having subtracted (2) from (1).}
 \end{figure}
 \subsection{The detail analysis of dI/dV curves}

 We analyze the dI/dV curves in detail. In the current formulation Eq.~(\ref{eq4}), the tunneling
 current is the sum of two parts J$_{1}$ and J$_{2}$. J$_{1}$ is proportional to the image part of
 the retarded Green function of the tip (or the density of state of the tip
 $\rho^{ct}(\omega)=-\frac{1}{\pi}ImG^{ct(r)}(\omega)$ ) and represents the tunneling
 from the surface and absorbed atom into the tip. J$_{2}$ is proportional to the density of
 electrons ($G_{ct}^{<} \propto n$) at the tip which represents the tunneling from the tip to the
 surface and absorbed atom. Experimentally, the shape of Kondo resonance in dI/dV curve, which is
 the mixing of Kondo resonance and others zero-bias structures, is not exactly the shape of
 Kondo resonance in the spectral function or density of state just like we have already discussed
 in the section B. The kondo resonance generally is modulated by a dip structure near
 zero-bias.

 We omit the Kondo Peak term Eq.~(\ref{eq8}) in the free Green function of the magnetic atom
 Eq.~(\ref{eq7}) and calculate the dI/dV curve again. We can find that from Fig.~\ref{fig7},
 the dI/dV curve is a dip. There is also a dip for empty surface when we remove the magnetic
 atom away from the surface. The dip generally becomes flat as increasing temperature.
 The behavior is similar to another important zero-bias phenomenon besides the Kondo
 resonance, the so-called zero-bias anomaly.\cite{Rowell1} If we subtract the dip from total
 dI/dV curve we get the Kondo resonance Peak (solid-line).

 The orthogonality catastrophe~\cite{Anderson2} of surface electrons prevents some
 many-body electronic transmission process. From the dI/dV curve we can find a local minimum near
 zero-bias, this is so-called zero-bias anomaly. The cotunneling mechanism has also contributions
 to the zero-bias anomaly even for nonmagnetic atom or quantum dot.\cite{Franceschi1,Weymann1}
 In this work the anomaly dip doesn't come from the orthogonality catastrophe because there
 aren't directly electronic interactions for the surface and tip. However the orthogonality
 catastrophe will contribute part of height of tunneling barrier from the surface to tip.
 The co-tunneling has been included in our model and self-consistent calculations.
 From Fig.~\ref{fig1} and the Feynman diagram in Fig.~\ref{fig2} for the Dyson equations, we can see
 that physics of co-tunneling is mainly included in the $\Sigma_{3}$ self-energies for the tip,
 surface and magnetic atom, such as an electron jumps to spin-up state under the Fermi energy after
 a spin-up electron having already jumped to the surface. In order to prominently illustrate the
 effects of co-tunneling, we drop the $\Sigma_{3}$ self-energies from the Dyson equations and
 calculate the dI/dV curve again. We find from Fig.~\ref{fig7} that the differential conductance
 decreases slightly. Remain dip structure is induced by subsequent single tunneling and higher
 power co-tunneling process embodied in the iterated calculations of the Dyson equations.
 The co-tunneling events do happen with short time interval and they only play the role of virtually
 intermediate states in perturbed calculations, but still have important influence on real
 physical tunneling.

 In the absence of the Kondo resonance, no dip structure near Fermi energies in the densities of state
 of the tip, magnetic atom and the surface show that the dip in dI/dV curve is dynamically tunneling
 phenomenon. Our results also indicate that the contributions of co-tunneling to zero-bias anomaly are
 not too large, but others tunneling events such as the subsequent single tunneling and the higher order
 co-tunneling contribute most parts of zero-bias anomaly. When Kondo resonance happens in the present
 of magnetic impurity, there is a dip structure in the density of state of surface conducting electron
 [Fig.\ref{fig3} (b)]. This contributes an additional dip structure to dI/dV curve, which mixes with
 the dip structure coming from dynamically tunneling. The dip structures have the widths with the same
 energy scale as the system temperature. If the Kondo temperature is higher than system temperature,
 the Kondo resonance is still modulated by the remain weaker dip structure [see Fig.\ref{fig7}] and
 the Kondo resonance at first sight splits two peaks. The total shape of dI/dV curve is the mixing of
 two splitting peaks and main strong dip structure.
  \begin{figure}\scalebox{0.8}{\includegraphics{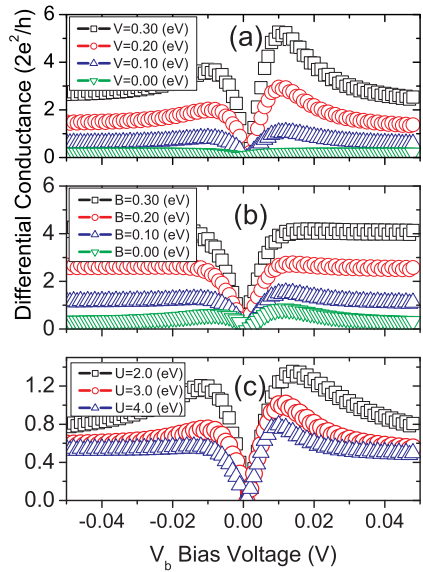}}
 \caption{\label{fig8}The dI/dV curves change with (a) the
 amplitudes V from magnetic atom to tip, (b) B from surface to tip
 and (c) on-site Coulomb energy U at 40 K}
 \end{figure}
 \subsection{Influence of tunneling amplitudes and electronic correlations}

 The change of the tunneling current of STM is very significant when we change the distance from
 the tip to surface. The tunneling matrix V or B represents the tunneling from the absorbed atom
 or the surface to the tip. V and B are proportional to $e^{-d(\vec r)/\lambda}$, where $d(\vec r)$
 is the distance from the tip to point $\vec r$ on the surface. The smaller is $d$ the larger is
 V or B. Thus, when the tip continuously accesses the surface or adsorbed atom, the parameter V or B
 changes from smaller value to larger one.  Fig.~\ref{fig8}(a) illustrates the resonance peaks in dI/dV
 curves become stronger and asymmetric when V increases from small to large values. In calculations,
 we shift the Kondo resonance about $\delta E$ = 0.005 eV above Fermi energy. We can also see that
 the Kondo resonance appears more prominent at strong tunneling in dI/dV curves.
 From Fig.~\ref{fig8}(b) we also see that strong tunneling from surface to tip makes the shape of
 dI/dV curves are more symmetric. This is because, at strong tunneling from the surface to tip,
 the role of the absorbed magnetic atom becomes less important. The electron-correlation interaction
 plays an important role for Kondo effect. Based on the Yosida and Yamada's
 calculations~\cite{Yosida1,Yamada1}, for strong coulomb interaction (large U) the Kondo resonance
 sharp but weak with small area closed by the energy-axis. After having calculated the dI/dV curves,
 we find the Kondo resonance more peak alike but weak as having increased U from 2.0 eV to 4.0 eV.

\section{\label{sect5} Summary and Conclusion}

 The STM tunneling spectra are obtained using the non-equilibrium self-consistent calculations.
 The dI/dV curves have Fano's shapes and are close to the experimental tunneling spectra of 3d
 transition-metal atoms on novel-metal surface. The Kondo resonance in measurements
 dI/dV curves generally are the mixture of pure Kondo resonance and the dip structure
 due to others zero-bias anomaly such as the subsequent single tunneling, cotunneling and other more
 complex tunneling events in the STM device. If Kondo resonance is strong the dI/dV curve is peak
 near zero-bias voltage and its shape is modulated by a valley or dip. Based on our model, the
 energy-window of tunneling electron has significant influence on the tunneling spectrum at very
 near zero-bias. The energy-window effect is unimportant for large bias voltage.

 On the one hand, Anderson model cann't capture all information in detail on the STM device,
 but it still captures the most important properties of STM device. Our calculations only provide
 the main characters of Kondo resonance in STM device.  More accurate calculations need to find
 a model more close to the real STM experiments and know the exact information on surface state.

 In this paper we have also developed a computational method of non-equilibrium electronic transport
 which is not only fit for STM device but also quantum dot, nano-molecule-device, hybridization
 system of normal metal and superconductor and others mesoscopic transport system. The self-consistent
 method can be extended to the system far beyond equilibrium by replace single-time integral and
 single-frequency Fourier transformation with double-times integral and double-frequency Fourier
 transformation. The method is suitable to the real time and real-space to study the pattern of
 tunneling spectrum on a surface.

\section*{Acknowledgements}

 The author are greatly indebted to Prof. Z. Zeng, Dr J. L. Wang and X. Q. Shi for their helpful
 suggestions and Dr. X. H. Zheng for having carefully read the first version of the manuscript.
 This work is financially supported by Nature Science Foundation of China and
 Knowledge Innovation Program of Chinese Academy of Sciences under
 KJCX2-SW-W11. This work had been partially running on the computers located at Center for
 computational science, Hefei Institutes of Physical Science, Chinese Academy of sciences.

\appendix
\section{Non-Equilibrium Green Function and Schwinger-Keldysh Contour}

 Schwinger and Keldysh~\cite{Schwinger1,Keldysh1} introduced a time
 contour or closed time path shown in the schematic diagram below which includes a positive
 branch $c_{+}$ from $-\infty$ to $+\infty$ and a negative branch $c_{-}$from $+\infty$ to
 $-\infty$.

 \begin{center}
 \setlength{\unitlength}{1cm}
 \begin{picture}(6,1)
 \put(6,0.5){\oval(0.4,0.4)[r]}
 \put(1,0.7){\vector(1,0){5}}
 \put(6,0.3){\vector(-1,0){5}}
 \put(1,0.85){-$\infty$}
 \put(1,0.05){-$\infty$}
 \put(6.5,0.5){+$\infty$}
 \put(3.5,0.85){$c^{+}$}
 \put(3.5,0.00){$c^{-}$}
 \end{picture}
 \end{center}

 A new Green function is introduced on the contour, which is expressed as
 $   G^{c}(t,t')=-i<T_{c}A(t)^{\dagger}B(t')>$, $T_{c}$ is an operator ordering the
 operators along the contour. If both $t$ and $t'$ are on the positive branch of the contour,
 $T_{c}$ is equivalent to the usual time-ordering operator and the Green function is the usual
 time Green function $   G^{\tau}(t,t')=-i<TA(t)^{\dagger}B(t')>$.
 If both $t$ and $t'$ are on the negative branch of the contour, $T_{c}$
 is expressed as $\tilde{T}$ which orders operators from $+\infty$ to $-\infty$. The Green
 function expresses as $   G^{\tilde{\tau}}(t,t')=-i<\tilde{T}A(t)^{\dagger}B(t')>$.
 If $t$ is on the positive branch and $t'$ is on the negative branch, the order of operators is
 well organized and $T_{c}$ is omitted, the Green function
 $   G^{<}(t,t') = i<A(t')^{\dagger}B(t)>$ is the lesser Green function.
 If $t$ is on the negative branch and $t'$ is on the positive
 branch, the order of operator is converted and $T_{c}$ is omitted, the Green function is the
 greater Green function and is written as $   G^{>}(t,t')=-i<A(t)B(t')^{\dagger}>$.

 By introducing the time contour, Feynman rules, Wick theorem and the Dyson equation are the same
 as that at zero temperature. The only difference is that the time integration now is along the
 contour and the time can be at different branches of the contour. The Dyson equation can be
 projected on the positive branch ($-\infty$,$+\infty$). If considering $t$ and $t'$ probably at
 different branches of the contour, there are four Dyson equations which can regroup into a
 matrix equation~\cite{Mahan1}
 \begin{widetext}
 \begin{eqnarray}
 \label{eq1a}
 G^{c}= \left(
   \begin{array}{ll}
         G^{\tau} & -G^{<} \\ G^{>} & -G^{\tilde{\tau}}
   \end{array}
   \right),~~
   \Sigma^{c}= \left(
   \begin{array}{ll}
        \Sigma^{\tau} &    -\Sigma^{<} \\    \Sigma^{>} &    -\Sigma^{\tilde{\tau}}
   \end{array}
   \right),~~
   g^{c} = \left(
   \begin{array}{ll}
    g^{\tau} &    -g^{<} \\    g^{>} &    -g^{\tilde{\tau}}
   \end{array}
   \right).
 \end{eqnarray}

 \begin{eqnarray}
 \label{eq2a}
 G^{c}(t,t') = g^{c}(t,t')
 + \int_{c^{+}}dt_{1}\int_{c^{+}}dt_{2}g^{c}(t,t_{1})\Sigma^{c}(t_{1},t_{2})G^{2}(t_{2},t').
 \end{eqnarray}
 \end{widetext}

 Now the contour Green function can be calculated in the same manner as that at zero temperature.
 The only difference is that complex Green functions and self energies are replaced by the complex
 matrixes of Green functions and self energies. In this work, we mainly study the electronic transport
 at small bias.  We assume the STM work in the non-equilibrium stationary state.~\cite{Zhou1}
 On the other hand the
 matrix Green functions define on only the positive branch. Thus, there is still the time translation
 invariance after the system enters into the non-equilibrium stationary state. The matrix Green functions
 are dependent on the relative time interval $G^{c}(t-t')$.  In energy space, the Dyson equation is more
 simple under the non-equilibrium stationary state and written as

 \begin{eqnarray}
 \label{eq3a}
 G^{c}(\omega) = g^{c}(\omega)+g^{c}(\omega)\Sigma^{c}(\omega)G^{c}(\omega).
 \end{eqnarray}

 The retarded and advanced Green functions are calculated using

 \begin{eqnarray}
 \label{eq4a}
 \Sigma^{r}(\omega)&=& \Sigma^{\tau}(\omega)-\Sigma^{<}(\omega)~;
 ~\Sigma^{a}(\omega)=\Sigma^{\tau}(\omega)-\Sigma^{>}(\omega) \\ \nonumber
    G^{r}(\omega)&=& G^{\tau}(\omega)-G^{<}(\omega)
 =G^{>}(\omega)-G^{\tilde{\tau}(\omega)} \\ \nonumber
    G^{a}(\omega)&=& G^{\tau}(\omega)-G^{>}(\omega)
 =G^{<}(\omega)-G^{\tilde{\tau}}(\omega).
 \end{eqnarray}

 In our calculations we have used following important properties of the matrix Green
 functions which can be derived from the symmetry of non-equilibrium Green function
 $G^{<(>)*}(\omega)=-G^{<(>)}(\omega)$,
 $G^{*\tau}(\omega)=-G^{\tilde{\tau}}(\omega)$,
 $G^{*r}(\omega)=G^{a}(\omega)$ and the Kubo-Martin-Schwinger condition
 $G^{<}(\omega)=-e^{-\beta(\omega-\mu)}G^{>}(\omega)$ \cite{Zhou1}

 \begin{eqnarray}
  \label{eq5a}
     G^{<}(\omega)&=&-2in_{f}(\omega)ImG^{r}(\omega),\\ \nonumber
     G^{>}(\omega)&=&-2i(n_{f}(\omega)-1)ImG^{r}(\omega) \\ \nonumber
     G^{\tau}(\omega)&=&
    ReG^{r}(\omega)-i(2n_{f}(\omega)-1)ImG^{r}(\omega), \\ \nonumber
     G^{\tilde{\tau}}(\omega)&=&
    -ReG^{r}(\omega)-i(2n_{f}(\omega)-1)ImG^{r}(\omega).
 \end{eqnarray}.
 \noindent where $n_{_{f}}(\omega)$ is the Fermi distribution function.

 \section{Derivation of the tunneling current formulation}

 It is convenient to calculate the contour Green function directly, not the lesser Green
 function. The mixing Green function in the current formulism (Eq.\ref{eq3}) can be defined on the contour
 and the full mixing contour Green function can be expressed using the Green functions of the
 tip, the adsorbed atom and the surface based on the methods in the reference.\cite{Jauho1}
 \begin{widetext}
 \begin{eqnarray}
  \label{eq6a}
  \langle t_{_{n\sigma}}^{\dagger}(t')d_{_{\sigma'}}(t) \rangle_{c}
  &=&\frac{1}{\hbar} \sum_{n'ss'}\int_{c}dt_{0}G_{nn's'\sigma}(t_{0},t')V_{n'ss'}^{*}
  g_{_{s\sigma'}}^{cd}(t_{0},t) \\ \nonumber
  \langle t_{_{n\sigma}}^{\dagger}(t')p_{_{k\sigma'}}(t) \rangle_{c}
  &=& \frac{1}{\hbar} \sum_{n'k'ss'}\int_{c}dt_{0}G_{nn's'\sigma}(t_{0},t')B_{n'k'ss'}^{*}
  g_{_{kk's\sigma'}}^{cp}(t_{0},t),
 \end{eqnarray}
 \end{widetext}

 \noindent $s$ and $s'$ are spin index. From the reference,\cite{Jauho1} if
 $A(t,t')=\int_{c}dt_{1}B(t,t_{1})C(t_{1},t')$ then

 \begin{eqnarray}
  \label{eq7a}
 A^{<}(t,t')=\int_{c^{+}}dt_{1}B^{r}(t,t_{1})C^{<}(t_{1},t')+B^{<}(t,t_{1})C^{a}(t_{1},t'),
 \end{eqnarray}

 After the contour Green function having projected onto the positive branch, the lesser Green
 function having inserted  into the current formulation and the current formulation having been
 reorganized, the current is written as $\langle J_{T} \rangle=J_{1}+J_{2}$, where
 \begin{eqnarray}
 \label{eq8a}
 J_{1} &=& \frac{2e}{\hbar^{2}}Re[\sum_{nn's\sigma}\int_{c^{+}} dt_{0}
            G_{nn's\sigma}^{ct(r)}(t_{0}-t')T_{nn's\sigma}^{ct(<)}(t_{0}-t)]
            \\ \nonumber
       &=& \frac{2e}{h}Re[\sum_{nn's\sigma}\int d\omega
            [G_{nn's\sigma}^{ct(r)}(\omega)T_{nn's\sigma}^{ct(<)}(\omega)]
            \\ \nonumber
       &=& \frac{2e}{h} \int d\omega ReTr[G^{ct(r)}(\omega)T^{ct(<)}(\omega)]
            \\ \nonumber
 J_{2} &=& \frac{2e}{\hbar^{2}}Re[\sum_{nn's\sigma}\int_{c^{+}} dt_{0}
            G_{nn's\sigma}^{ct(<)}(t_{0}-t')T_{nn's\sigma}^{ct(a)}(t_{0}-t)]
            \\ \nonumber
       &=& \frac{2e}{h}Re[\sum_{nn's\sigma}\int d\omega
            G_{nn's\sigma}^{ct(<)}(\omega)T_{nn's\sigma}^{ct(a)}(\omega)]
            \\ \nonumber
       &=& \frac{2e}{h}\int d\omega ReTr[G^{ct(<)}(\omega)T^{ct(a)}(\omega)].
 \end{eqnarray}

 \noindent and the first equation converts to the second equation by using the Fourier
 transformations under the same assumption of non-equilibrium stationary state
 as the Dyson equations Eq.(~\ref{eq3a}).   The T matrixes are written as,
 \begin{eqnarray}
 \label{eq9a}
      T_{nn's\sigma}^{ct(<)}(\omega) &=& \sum_{kk's'\sigma'}
                         B_{_{nk\sigma\sigma'}}B_{_{n'k'ss'}}^{*}g_{_{kk's'\sigma'}}^{cp(<)}(\omega)
                       \\ \nonumber  &+& \sum_{s'\sigma'}
                         V_{_{n\sigma\sigma'}}V_{_{n'ss'}}^{*}g_{_{s'\sigma'}}^{cd(<)}(\omega)
                         \\ \nonumber
      T_{nn's\sigma}^{ct(a)}(\omega) &=& \sum_{kk's'\sigma'}
                         B_{_{nk\sigma\sigma'}}B_{_{n'k'ss'}}^{*}g_{_{kk's'\sigma'}}^{cp(a)}(\omega)
                  \\ \nonumber  &+& \sum_{s'\sigma'}
                         V_{_{n\sigma\sigma'}}V_{_{n'ss'}}^{*}g_{_{s'\sigma'}}^{cd(a)}(\omega).
 \end{eqnarray}
 \noindent $g_{_{kk's'\sigma'}}^{cp(<)}$ and $g_{_{s'\sigma'}}^{cd(<)}$ are the free Green functions
 of the absorbed atom and the surface. In order to calculate contour Green function, at first,
 the self-energy is calculated. Based on Feynman diagrams [Fig.\ref{fig2}], the self energies of
 the absorbed atom and the surface can be calculated using the following formulations:

 \begin{eqnarray}
 \label{eq10a}
       \Sigma_{\alpha\beta}^{cd}(\omega) &=&
                       \sum_{n'n''\sigma'\sigma''}
                         V_{_{n'\sigma'\alpha}}V_{_{n''\sigma''\beta}}^{*}g_{_{n'n''\sigma'\sigma''}}^{ct}(\omega)
                         \\ \nonumber &+&
                       \sum_{k'k''\sigma'\sigma''}
                         A_{_{k'\sigma'\alpha}}A_{_{k''\sigma''\beta}}^{*}g_{_{k'k''\sigma'\sigma''}}^{cp}(\omega)
                          + \Sigma_{3\alpha\beta}^{cd}(\omega) \\ \nonumber
      \Sigma_{kl\alpha\beta}^{cp}(\omega) &=&
                       \sum_{n'n''\sigma'\sigma''}
                         B_{_{n'\sigma'k\alpha}}B_{_{n''\sigma''l\beta}}^{*}g_{_{n'n''\sigma'\sigma''}}^{ct}(\omega)
                         \\ \nonumber &+&
                       \sum_{\sigma'\sigma''}
              A_{_{k\alpha\sigma'}}A_{_{l\beta\sigma''}}^{*}g_{_{\sigma'\sigma''}}^{cd}(\omega)
               + \Sigma_{3kl\alpha\beta}^{cp}(\omega) \\ \nonumber
    \Sigma_{mn\alpha\beta}^{ct}(\omega) &=&
                       \sum_{\alpha'\beta'}
                         V_{_{m\alpha\alpha'}}V_{_{n\beta\beta'}}^{*}G_{_{\alpha'\beta'}}^{cd}(\omega)
                       \\ \nonumber &+&
                       \sum_{k'l'\alpha'\beta'}
                         B_{_{mk'\alpha\alpha'}}B_{_{nl'\beta\beta'}}^{*}G_{_{k'l'\alpha'\beta'}}^{cp}(\omega)
                       + \Sigma_{3mn\alpha\beta}^{ct}(\omega),
 \end{eqnarray}
 \noindent $\alpha$ and $\beta$ are the spin index. The $\Sigma_{3}$ self energies for the
 tip, adatom and surface are expressed as
 \begin{widetext}
 \begin{eqnarray}
 \label{eq11a}
     \Sigma_{3kl\alpha\beta}^{cp}(\omega) &=&
                       \sum_{mn\alpha'\beta'\alpha''\beta''}
                         B_{_{mk\alpha\alpha'}}^{*}V_{_{n\beta'\alpha''}}^{*}A_{_{l\beta''\beta}}
                         g_{_{\alpha''\beta''}}^{cd}(\omega)g_{_{mn\alpha'\beta'}}^{ct}(\omega)
                         \\ \nonumber &+&
                       \sum_{mn\alpha'\beta'\alpha''\beta''}
                         A_{_{k\alpha\alpha'}}^{*}V_{_{m\beta'\alpha''}}B_{_{ml\beta''\beta}}
                         g_{_{mn\alpha''\beta''}}^{ct}(\omega)g_{_{\alpha'\beta'}}^{cd}(\omega)
                         \\ \nonumber
     \Sigma_{3mn\alpha\beta}^{ct}(\omega) &=&
                       \sum_{k'l'\alpha'\beta'\alpha''\beta''}
                         V_{_{m\alpha\alpha'}}^{*}A_{_{k'\beta'\alpha''}}B_{_{nl'\beta''\beta}}^{*}
                         g_{_{\alpha'\beta'}}^{cd}(\omega)g_{_{k'l'\alpha''\beta''}}^{cp}(\omega)
                         \\ \nonumber &+&
                       \sum_{k'l'\alpha'\beta'\alpha''\beta''}
                         B_{_{mk'\alpha\alpha'}}A_{_{l'\beta'\alpha''}}^{*}V_{_{n\beta''\beta'}}
                         g_{_{k'l'\alpha'\beta'}}^{cp}(\omega)g_{_{\alpha''\beta''}}^{cd}(\omega)
                         \\ \nonumber
      \Sigma_{3\alpha\beta}^{cd}(\omega) &=&
                       \sum_{mnk'l'\alpha'\beta'\alpha''\beta''}
                         A_{_{k'\alpha\alpha'}}B_{_{ml'\beta'\alpha''}}^{*}V_{_{n\beta''\beta}}^{*}
                         g_{_{k'l'\alpha'\beta'}}^{cp}(\omega)g_{_{mn\alpha''\beta''}}^{ct}(\omega)
                         \\ \nonumber &+&
                       \sum_{k'l'\alpha'\beta'\alpha''\beta''}
                         V_{_{m\alpha\alpha'}}B_{_{nk'\beta'\alpha''}}A_{_{l'\beta''\beta}}^{*}
                         g_{_{mn\alpha'\beta'}}^{ct}(\omega)g_{_{k'l'\alpha''\beta''}}^{cp}(\omega).
 \end{eqnarray}
 \end{widetext}

 \noindent which presents the cotunneling events in the STM device. A typical cotunneling event is
 that an electron jumps to the surface and leaved a hole on the magnetic atom, and subsequently a tip
 electron fills the hole on the magnetic atom [Fig.~\ref{fig1} and Fig.~\ref{fig2}].
 For higher-order cotunneling processes
 include more electrons and tunneling events which generally are included in the iterated calculations.
 By compared the matrix expression of self-energy with the matrix expression of T matrix in the current
 formulation, $T$ is just the lowest order approximation of the exact self-energy. In our current
 formulation $T$ replaces by the self-energy $\Sigma$ in Eq.~(\ref{eq8a}).
 \begin{eqnarray}
 \label{eq12a}
    J_{1} &=& \frac{-2e}{h}\int_{-\infty}^{+\infty} d\omega ReTr[G^{ct(r)}(\omega)\Sigma^{ct(<)}(\omega)]
            \\ \nonumber
    J_{2} &=& \frac{-2e}{h}\int_{-\infty}^{+\infty} d\omega ReTr[G^{ct(<)}(\omega)\Sigma^{ct(a)}(\omega)].
 \end{eqnarray}
 When near zero bias the electronic tunneling is in classic forbidden
 regime, we use following current formula with energy cutoffs

 \begin{eqnarray}
 \label{eq12b}
 \langle J_{T} \rangle &=& J_{1}+J_{2} \\ \nonumber
       J_{1} &=& \frac{-2e}{h}\int_{\mu}^{\mu+eV_{b}} d\omega
                            Re Tr[G^{ct(r)}(\omega)\Sigma^{ct(<)}(\omega)] \\ \nonumber
       J_{2} &=& \frac{-2e}{h}\int_{\mu}^{\mu+eV_{b}} d\omega
                            Re Tr[G^{ct(<)}(\omega)\Sigma^{ct(a)}(\omega)].
 \end{eqnarray}
 \noindent where if positive bias voltage applies to surface $V_{b}>0$.
  $\mu$ is equal to $\mu_{S}$ the chemical potential
 of the surface. Because of numerical reason $\mu_{S}$ sets to zero value to keep the
 position of Kondo peak unchange. The bias-voltage is applied by changing the
 chemical potential of the tip in our numerical calculations. $\mu_{S}$ keeps zero
 when applying a small bias voltage V$_{b}$ on the surface mean that applying
 a bias voltage to the tip V$_{t}$=-V$_{b}$.

 In real calculations, we add a minus ahead the current
 formulas Eq.(\ref{eq4}), Eq.(\ref{eq12a}) and Eq.(\ref{eq12b})) to make the current
 $J_{T}=J_{1}+J_{2} >0$ at positive bias once the lower bound and up-bound
 of the integrates are clearly written in the current formulas.
 We can adopt the chemical potential langue just like in
 Ref~\cite{Wingreen1} without introducing additional minus in the current formulas.
 The basic reason of the difference is that the increase of
 chemical potential $\delta\mu$ of electron is equal to -e$\delta V_{b}$


\begin{references}

 \bibitem{Madhavan2}V. Madhavan, W. Chen, T. Jamneada, M. F. Crommie, and N. S. Wingreen,
  Science {\bf 280}, 567 (1998).
 \bibitem{Nagaoka1} K. Nagaoka, T. Jamneada, M. Grobis, and M. F. Crommie,
   Phys. Rev. Lett. {\bf 88}, 077205 (2002).
 \bibitem{Knorr1} N. Knorr, M. A. Schneider, Lars Dickh\"oner, P. Wahl, and K. Kern,
   Phys. Rev. Lett. {\bf 88}, 096804 (2002).
  \bibitem{Schneider1}  M. A. Schneider, L. Vitali, N. Knorr, and K. Kern,
   Phys. Rev. B {\bf 65}, R121406 (2002).
 \bibitem{Jamneala1} T. Jamneala, V. Madhavan, W. Chen, and M. F. Crommie,
  Phys. Rev. B {\bf 61}, 9990 (2000).
 \bibitem{Li1} Jiutao Li, Wolf-Dieter Schneider, Richard Berndt,
 Bernard Delley, Phys. Rev. Lett. {\bf 80}, 2893 (1998).
 \bibitem{Zhao1} Aidi Zhao, Qunxiang Li, Lan Chen, Hongjun Xiang,
  Weihua Wang, Shuan Pan, Bing Wang, Xudong Xiao, Jinlong Yang, J.
  G. Hou, Qingshi Zhu, Science {\bf 309}, 1542 (2005).
 \bibitem{Wahl1} P. Wahl, L. Diekh\"oner, G. Wittich, L. Vitali,
   M. A. Schneider, and Kern, Phys. Rev. Lett. {\bf 95}, 166601 (2005).
 \bibitem{Madhavan1} V. Madhavan, T. Jamneada, K. Nagaoka,
  W. Chen, Je-Luen Li, Steven G. Louie, and M. F. Crommie,
  Phys. Rev. B {\bf 66}, 212411 (2002).
 \bibitem{Kudasov1} Y. B. Kudasov, and V. M. Uzdin,
   Phys. Rev. Lett. {\bf 89}, 276802 (2002).
 \bibitem{Manoharan1} H. C. Manoharan, C. P. Lutz, and D. M. Eigler,
  Nature(London) {\bf 403}, 512 (2000).
 \bibitem{Fiete1} G. A. Fiete, and E. J. Heller, Rev. Mod. Phys. {\bf 75}, 933 (2003).
 \bibitem{Tersoff1} J. Tersoff, and D. R. Hamann,
  Phys. Rev. Lett. {\bf 50}, 1998 (1983).
 \bibitem{Tersoff2} J. Tersoff, and D. R. Hamann,
  Phys. Rev. B {\bf 31}, 805 (1985).
 \bibitem{Bardeen1} J. Bardeen,
  Phys. Rev. Lett. {\bf 6}, 57 (1961).
 \bibitem{Bracher1} C. Bracher, M. Riza, and M. Kleber
   Phys. Rev. B {\bf 56}, 7704 (1997).
 \bibitem{Schiller1} A. Schiller, and S. Hershfield,
   Phys. Rev. B {\bf 61}, 9036 (2000).
 \bibitem{Lin1} Chiung-Yuan Lin, A. H. Castro Neto, and B. A. Jones,
  arXiv:cond-mat/0307185 (2003).
 \bibitem{Luo1} H. G. Luo, T. Xiang, X. Q. Wang, Z. B. Su, and L. Yu, Phys. Rev. Lett.
 {\bf 92}, 256602 (2004).
 \bibitem{Ujsaghy1} O. \'Ujs\'aghy, J. Kroha, L. Szunyogh, and A. Zawadowski, Phys. Rev. Lett.
 {\bf 85}, 2557 (2000).
 \bibitem{Merino1} J. Merino, and O. Gunnarsson, Phys. Rev. B {\bf 69}, 115404 (2004).
 \bibitem{Plihal1} M. Plihal, and J. W. Gadzuk,  Phys. Rev. B {\bf 63}, 085404 (2001).
 \bibitem{Hewson1} A.C. Hewson, The Kondo Problem to Heavy Fermions, Cambridge University Press (1993).
 \bibitem{Coleman1} Piers Coleman, Phys. Rev. B {\bf 29}, 3035 (1983).
 \bibitem{Bickers1} N.E. Bickers, Reviews of Modern Physics, {\bf 59}, 845 (1987).
 \bibitem{Schonhammer1} K. Sch\"onhammer, Solid State Communications, {\bf 22}, 51 (1977).
 \bibitem{Kajueter1} H. Kajueter, and G. Kotliar, Phys. Rev. Lett. {\bf 77}, 131 (1996).
 \bibitem{Gunnarsson1} O. Gunnarsson, and K. Sch\"onhammer, Phys. Rev. B {\bf 28}, 4315 (1983).
 \bibitem{Anderson2} P. W. Anderson, Phys. Rev. Lett. {\bf 18}, 1049 (1967).
 \bibitem{Franceschi1} S. De Franceschi, S. Sasaki, J.M. Elzerman, W.G. van der Wiel, S. Tarucha,
                       and L.P. Kouwenhoven, Phys. Rev. Lett. {\bf 86}, 878 (2003).
 \bibitem{Weymann1} I. Weymann, J. Barna\'s, J. K\"onig, J. Martinek, and G. Sch\"on,
                    arXiv:cond-mat/0412434, (2004).
 \bibitem{Schwinger1} J. Schwinger, {\bf } J. Math. Phys. {\bf 2}, 407 (1961).
 \bibitem{Keldysh1} L. V. Keldysh, Zh. Eksp. Teor. Fiz
  {\bf 47}, 1515 (1964); [Sov. Phys. JETP {\bf 20}, 1018 (1965)]
 \bibitem{Mahan1} Gerald D. Mahan Many-Particle Physics (Second Edition) chapter 2, Plenum Press (1990);
 \bibitem{Zhou1} Guang-zhao Zhou (K.C. Chou), Z.B. Su, B.L.Hao, and L. Yu, Physics Reports {\bf 118}, 1-131
 (1985); J. Rammer and H. Smith, Reviews of Modern Physics, {\bf 58}, 323 (1986).
 \bibitem{Kamenev1} Alex Kamenev,
   Lecture Notes Presented at Windsor NATO School on "Field Theory of Strongly Correlated Fermions and Bosons
   in Low-Dimensional Disordered System" (August 2001). arXiv:cond-mat/0109316 (2001).
 \bibitem{Caroli1} C. Caroli, R. Combescot, P. Nozieres, and D. Saint-James,
  J. Phys. C: Solid St. Phys. {\bf 4}, 916 (1971).
 \bibitem{Meir1} Yigal Meir, and N. S. Wingreen, Phys. Rev. Lett. {\bf 68}, 2512 (1992).
 \bibitem{Wingreen1} N. S. Wingreen, and Yigal Meir, Phys. Rev. B {\bf 49}, 11040 (1994).
 \bibitem{Jauho1} Antti-Pekla Jauho, N. S. Wingreen, and Yigal Meir, Phys. Rev. B {\bf 50}, 5528 (1994).
 \bibitem{Baigeng1} Baigeng Wang, Jian Wang, and Hong Guo, Phys. Rev. Lett. {\bf 82}, 398 (1999).
 \bibitem{Anantram1} M. P. Anantram, and S. Datta, Phys. Rev. B {\bf 51}, 7632 (1995).
 \bibitem{Haldane1} F. D. M. Haldane, Phys. Rev. Lett. {\bf 40}, 416 (1978).
 \bibitem{Wilson1} K. G. Wilson, Rev. Mod. Phys. {\bf 47}, 773 (1975).
 \bibitem{Schrieffer1} J. R. Schrieffer, and P. A. Wolff, Phys. Rev. {\bf 149}, 491 (1966).
 \bibitem{Yosida1} K. Yosida and K. Yamada, Supplement of Progress of Theoretical Physics, {\bf 46}, 244 (1970).
 \bibitem{Yamada1} K. Yamada, Progress of Theoretical Physics, {\bf 53}, 970 (1975).
 \bibitem{Kaga1} H. Kaga, H. Kube, and T. Fujiwara, Phys. Rev. B {\bf 37}, 341 (1988).
 \bibitem{Luttinger} J. M. Luttinger, Phys. Rev. {\bf 121}, 942 (1960).
 \bibitem{Agam1} Oded Agam, and Avraham Schiller, Phys. Rev. Lett. {\bf 86}, 484 (2001).
 \bibitem{Rowell1} J. M. Rowell and L. Y. L. Shen, Phys. Rev. Lett. {\bf 17}, 15 (1966).
\end{references}
\end{document}